\documentclass[runningheads]{llncs}
\usepackage[T1]{fontenc}
\usepackage{graphicx,verbatim}
\usepackage{booktabs}
\usepackage{float}
\usepackage{marvosym}
\usepackage{amsmath}
\usepackage{multirow}
\usepackage[caption=false]{subfig} 

\usepackage[colorlinks=true, allcolors=blue]{hyperref}
\begin{document}
%

\title{Fairness-Aware Data Augmentation for Cardiac MRI using Text-Conditioned Diffusion Models}

\titlerunning{Fairness-Aware Data Augmentation using Text-Conditioned Diffusion Models}
%

%
\author{
Grzegorz Skorupko \inst{1}$^{(\textrm{\Letter})}$ 
\and
Richard Osuala \inst{1,2,3} 
\and
Zuzanna Szafranowska\inst{1} 
\and
Kaisar Kushibar \inst{1} 
\and
Vien Ngoc Dang \inst{1} 
\and
Nay Aung \inst{4,5} 
\and
Steffen E. Petersen \inst{4,5} 
\and
Karim Lekadir \inst{1,6} 
\and
Polyxeni Gkontra\inst{1} 
}

\authorrunning{G. Skorupko et al.}

\institute{Barcelona Artificial Intelligence in Medicine Lab (BCN-AIM), Departament de Matemàtiques i Informàtica, Universitat de Barcelona, Spain
\email{grzegorz.skorupko@ub.edu} \and
Helmholtz Center Munich, Munich, Germany
\and
Technical University of Munich, Munich, Germany
\and
William Harvey Research Institute, NIHR Barts Biomedical Research Centre, Queen Mary University London, Charterhouse Square, London, UK \and
Barts Heart Centre, St Bartholomew’s Hospital, Barts Health NHS Trust, West Smithfield, London, UK \and
Institució Catalana de Recerca i Estudis Avançats (ICREA), Barcelona, Spain}




\maketitle              
\begin{abstract}
While deep learning holds great promise for disease diagnosis and prognosis in cardiac magnetic resonance imaging, its progress is often constrained by highly imbalanced and biased training datasets.
To address this issue, we propose a method to alleviate imbalances inherent in datasets through the generation of synthetic data based on sensitive attributes such as sex, age, body mass index (BMI), and health condition. We adopt ControlNet based on a denoising diffusion probabilistic model to condition on text assembled from patient metadata and cardiac geometry derived from segmentation masks. We assess our method using a large-cohort study from the UK Biobank  by evaluating the realism of the generated images using established quantitative metrics. Furthermore, we conduct a downstream classification task aimed at debiasing a classifier by rectifying imbalances within underrepresented groups through synthetically generated samples. Our experiments demonstrate the effectiveness of the proposed approach in mitigating dataset imbalances, such as the scarcity of diagnosed female patients or individuals with normal BMI level suffering from heart failure. This work represents a major step towards the adoption of synthetic data for the development of fair and generalizable models for medical classification tasks. Notably, we conduct all our experiments using a single, consumer-level GPU to highlight the feasibility of our approach within resource-constrained environments. Our code is available at
\url{https://github.com/faildeny/debiasing-cardiac-mri}.

\end{abstract}
\section{Introduction}

Cardiovascular diseases remain the main cause of mortality worldwide, accounting for approximately one third of annual deaths globally~\cite{jagannathan_global_2019}. Cardiovascular magnetic resonance (CMR) is currently the gold standard in evaluating the structure and function of the heart. However, its acquisition is expensive and the annotation process of multi-slice cine sequences requires a significant amount of time. Consequently, the amount of available training data is limited, hindering the adoption of deep learning based algorithms. Despite the efforts to automate CMR dataset collection, annotation and analysis, end-to-end models are still not common. Such solutions are more affected by the inherent biases in the training data especially when the data is scarce. For example, Puyol et al.~\cite{fairness_puyol} showed discrepancies in the performance of CMR segmentation models for subgroups based on sex and race. This finding was primarily attributed to the pronounced imbalance in the training dataset, which consisted mostly of individuals of white race. Such biases can significantly influence the decision-making process of classification models and were widely studied and addressed in various medical domains~\cite{biased_classifiers,biased_classifiers2,biased_classifiers_with_method,detect_and_correct_bias,detect_and_correct2}.

Advancements in generative deep learning models opened paths to previously unexplored approaches in tackling this crucial challenge in machine learning, namely, algorithmic bias. Some studies have proposed bias mitigation methods through different sampling strategies or modifications to model architecture and training procedures~\cite{bias_reduction_strategies,adversarial_learning}. Nonetheless, in the medical domain, the adoption of generative models to mitigate biases through the use of synthetic data has received relatively little attention. Recent works based on GANs
and Diffusion models
focusing on dermatology, chest X-ray and histopathology domains, are among the very few examples in this direction \cite{mikolajczyk_biasing_2022,ktena2023generative}. Ktena et al. \cite{ktena2023generative} proposed models conditioned on both diagnostic and sensitive attributes, such as sex, age, or skin tone, allowed to augment the unbalanced training dataset and successfully reduce the biases in classification tasks. However, to the best of our knowledge, none of the previous works focused on magnetic resonance imaging (MRI) or cardiovascular domain, nor did they allow for conditioning image generation on shape information from segmentation masks or textual prompts.

\begin{figure}[b]

    \includegraphics[width=\linewidth]{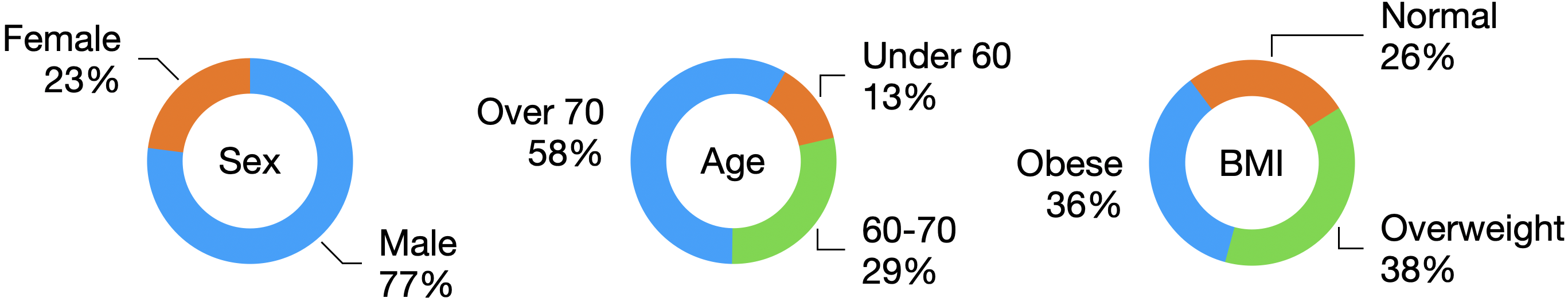}
    \caption{Demographic statistics of patients diagnosed with heart failure from the UK Biobank imaging study.}
    \label{fig:demographics}
\end{figure}

To address this gap, we propose an open-source pipeline involving training of a resource-intensive stable diffusion model~\cite{Rombach_2022_CVPR} within a limited computational environment. More precisely, we implement a latent diffusion model with combined text and image inputs to generate spatially consistent CMR frames to mitigate biases introduced by unbalanced training data in CMR-based deep learning models for disease diagnosis. This approach facilitates the generation of CMR data for underrepresented patient subgroups, considering factors such as sex, age, BMI, and heart conditions, alongside spatial-temporal features defined through segmentation masks from multiple cardiac phases. We evaluate the quality of the generated images using the domain-specific, recently introduced~\cite{osuala2024towards} and validated~\cite{konz2025frechetradiomicdistancefrd} Fréchet Radiomics Distance (FRD) score. Furthermore, we assess the impact of the attributed-conditioned synthetic images in heart failure classification model training, demonstrating enhanced model fairness and performance across diverse patient subgroups. Overall, the proposed approach serves as a general-purpose targeted augmentation method, as we illustrate its applicability in resource-limited environments. 
Our key contributions are:

\begin{enumerate}
\item A promising data augmentation method for improving fairness through an Attribute-Conditioned Latent Diffusion Model.
\item The first application of a diffusion model to explicitly address fairness in cardiac MRI, and the first to condition on BMI---an important but underexplored factor in this modality.
\item We experimentally demonstrate that the proposed method simultaneously improves both fairness and classification performance across subgroups, highlighting its potential for clinical adoption.
\end{enumerate}



\section{Methodology}
\subsection{Dataset}
For this study, we use the UK Biobank (UKBB)~\cite{sudlow2015uk}, a large-scale resource with data from over 500,000 participants recruited between 2006 and 2010, that includes demographics, electronic health records (EHRs), biomarkers, and genomics. We focus on a subset of patients who participated in the imaging study and underwent CMR scans. In total, our dataset consists of 25480 multi-slice, short-axis cine CMRs with annotations for end-diastole (ED) and end-systole (ES) frames. The annotation masks label key cardiac structures: left and right ventricles and myocardium. Based on International Classification of Diseases (ICD-10) codes from in-hospital patient data, we identified a subset of 270 patients diagnosed with heart failure at the time of the CMR acquisition.  Fig.~\ref{fig:demographics} provides the distribution of characteristics of the participants included in the study. In our analysis, we divided patients into groups by age: below 60, 60-70 and over 70 years old, by BMI: below 25 (underweight and normal), 25-30 (overweight) and over 30 (obese), and by sex.

\subsubsection{Data pre-processing}
Due to the multidimensional nature of CMR samples (4D), we conduct several data preprocessing steps to adapt to the image format most commonly used in state-of-the-art classification models, i.e. 2D, 3-channel images, to be generated by the Stable diffusion model with ControlNet. We extract the central slice from each volume and stack cine frames from ED and ES phases as color channels, creating a 2D RGB image. To keep the advantage of multidimensional data, we extract three central slices per patient and include cine frames before and after ED and ES, increasing training images nine-fold. It should be noted that we do not apply this augmentation to the validation or test sets, where we solely use one central slice with ED and ES frames.

\subsection{Conditioned image generation}
\begin{figure}[b]
    \centering
    \includegraphics[width=\textwidth]{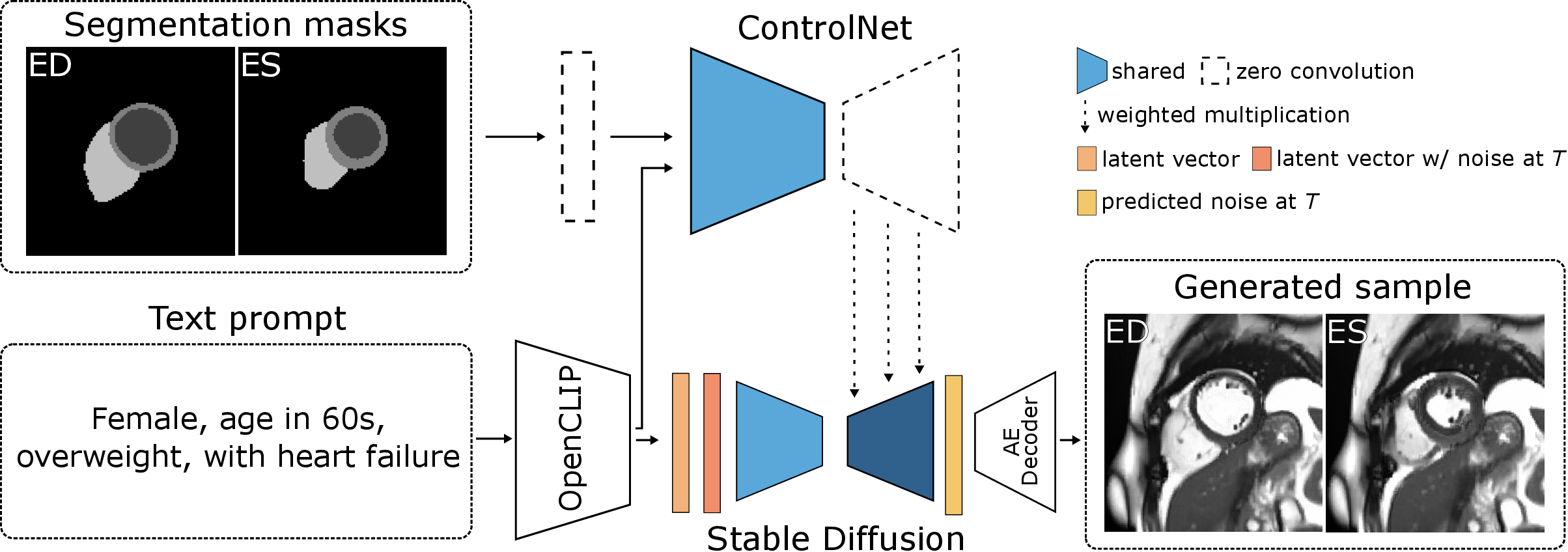}
    \caption{Overview of the proposed pipeline for generating synthetic CMR data conditioned on textual information and cardiac geometry derived from segmentation masks.}
    \label{fig:controlnet_diagram}
\end{figure}

An overview of the proposed pipeline for generating synthetic CMR images based on textual information and cardiac masks is provided in Fig.~\ref{fig:controlnet_diagram}. We use ControlNet~\cite{zhang2023adding}, which enhances Stable Diffusion~\cite{Rombach_2022_CVPR} by enabling fine-tuning with text and image inputs. The approach duplicates the pretrained model, adding spatial input only to the cloned branch, which connects to the original architecture via zero convolution layers to reduce noise and preserve the trainable copy’s backbone. The original model’s weights remain locked to retain generative capabilities, allowing adaptation to new imaging domains without costly retraining.

\subsubsection{Diffusion model training}
We conduct all experiments on a single Nvidia 3080Ti GPU with 16GB of memory.
To train the diffusion model, we adopt the implementation provided by~\cite{zhang2023adding}.  To fully leverage the advantages of the pretrained model, we upscale the training samples to 512x512 pixels to match the final pretraining resolution of a Stable diffusion 2.1-base model~\cite{Rombach_2022_CVPR}.
In the training setup, we use the pretrained image AutoEncoder network and the OpenClip~\cite{ilharco_gabriel_2021_5143773} text encoder pretrained on the LAION-5B~\cite{schuhmann2022laionb} dataset. During the training phase, we exclusively fine-tune the ControlNet branch of the model. In this setup, it is possible to train the model with batch size of 1 with 2 gradient accumulations. We train the model with a learning rate of 1e-5 for 5 epochs, which takes approximately 3 days in our setup. All the code is based on PyTorch framework~\cite{pytorch}.

\subsubsection{Debiased dataset generation}
To address biases resulting from underrepresentation of certain groups, we use weighted random sampling on our initial dataset. Patients are grouped based on sex, age, BMI and diagnosis, which creates 36 groups in total. For example, female, overweight patients younger than 60 years that are healthy belong to the same subset. Based on each group's population we calculate the sampling weights that are inversely proportional to their size. We subsequently generate synthetic images based on existing prompts and masks for underrepresented groups. This way, we ensure that imaging inputs are coherent with patient's characteristics and do not contribute to additional noise.

\subsection{Downstream classification model training}
For the downstream classification task, due to the relatively small dataset size, we use a well established ResNet-18 model~\cite{ResNet} with weights pretrained on the ImageNet dataset~\cite{ImageNet}. All training samples are scaled to the native pretraining resolution of 224x224 pixels. Models are trained for 10 epochs with a batch size of 64, starting at a 1e-4 learning rate, reduced by 2 on plateau for 3 epochs. Standard augmentations like random flipping and Gaussian noise are applied. We save model weights after each epoch, selecting the best checkpoint based on balanced accuracy. The dataset is split into 20\% test data, with 20\% of the training set reserved for validation.
During training, we explore different sampling methods, including sample weighting (SW), which adjusts weights based solely on label, and stratified sample weighting (SSW), which considers 
the joint distribution of subject label and sensitive subgroup.

\subsection{Evaluation metrics}

\subsubsection{Synthetic data evaluation}

To evaluate synthetic medical image quality, we use the radiology domain-specific FRD, thereby avoiding the limitations of alternatives such as the Fréchet Inception Distance (FID)\cite{fid}, which, pretrained on natural images, often lacks robustness in medical imaging~\cite{xing2023you,konz2025frechetradiomicdistancefrd}.
In contrast, FRD measures distances between distributions of radiomics features, which are a proven method for characterizing medical images \cite{osuala2024towards,konz2025frechetradiomicdistancefrd,szabo_radiomics_2023,lambin2012radiomics}. 
To assess the model’s ability to condition images on sensitive attributes, we compute FRD within subpopulations (e.g., only females) and between groups (e.g., females vs males). This allows us to evaluate how well real image feature distributions are preserved in data generated by our model.
\subsubsection{Classification task}


To evaluate classifier performance on heart failure diagnosis, we use AUROC and Balanced Accuracy (BACC), the latter addressing class imbalance due to disease prevalence of {\raise.17ex\hbox{$\scriptstyle\sim$}}1\%. Metrics are reported globally, per subgroup, and as an average between groups to ensure equal importance across populations, providing a fairer assessment of model performance.



For fairness evaluation, we use the Equal Opportunity Difference (EOD). EOD measures the disparity in true positive rates (TPR) across different demographic groups, ensuring that the model performs equally well for all groups in terms of correctly identifying positive outcomes. 
The formula for EOD is given by:
\begin{equation}
\text{EOD} = \min_{x \in \Omega_X} \text{TPR}_{x} - \max_{x \in \Omega_X} \text{TPR}_{x},
\end{equation}
where \(\text{TPR}_{x}\) represents the true positive rate for group \(x\), and \(\Omega_X\) denotes the set of all groups under consideration.

\section{Results}
\subsection{Synthetic data evaluation} \label{synth_evaluation}
\subsubsection{FRD scores comparison between subpopulations}
\begin{figure}[t]
    \centering
    \includegraphics[width=\textwidth]{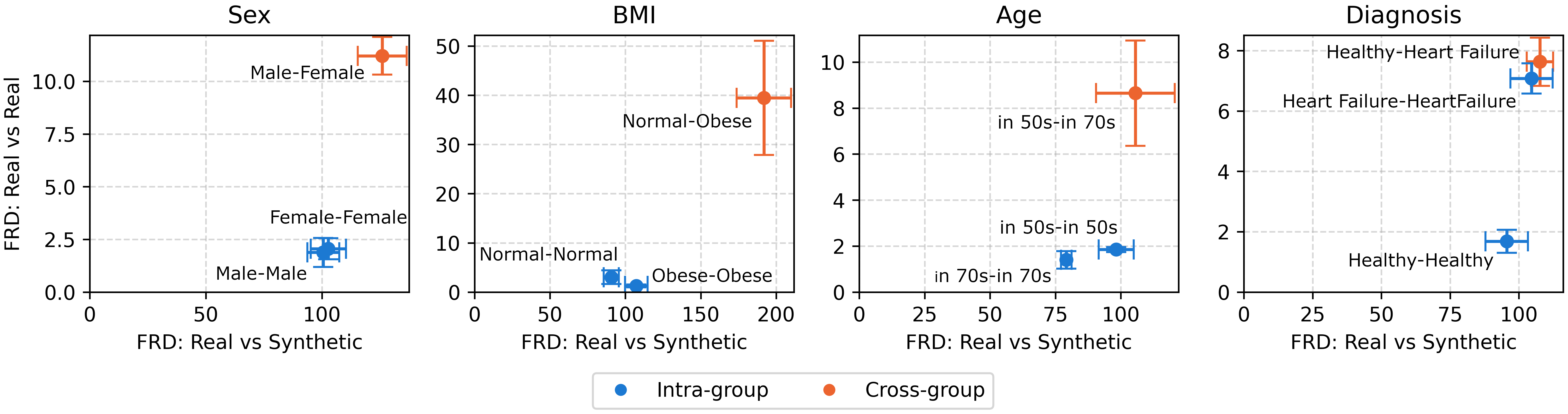}
    \caption{FRD scores within original dataset (Real vs Real) and with synthetic data (Real vs Synthetic) calculated for attribute subpopulations. }
    \label{fig:frd_plot}

\end{figure}

Fig.~\ref{fig:frd_plot} shows FRD values for CMR images across subgroups categorized by sex, age, BMI, and health condition. The vertical axis (Real vs. Real) captures visual differences in real data, while the horizontal axis (Real vs. Synthetic) evaluates how well these differences are preserved in synthetic images. Intra-group comparisons (Female-Female) yield lower FRD scores, while cross-group (Male-Female) show higher values, indicating expected dissimilarities. Synthetic images have higher FRD scores but follow a similar trend. A comparable pattern appears for BMI, where synthetic images of obese patients closely resemble real high-BMI subjects. For age, real datasets show notable radiomics feature differences, which are less distinct in synthetic images, especially for younger patients. Finally, differences between heart failure and healthy individuals are subtler than for other attributes in both real and synthetic data, highlighting the difficulty of the diagnosis task.

\subsubsection{Qualitative analysis}
\begin{figure}[tb]

    \centering
    \subfloat[Underweight]{\includegraphics[width=0.20\linewidth]{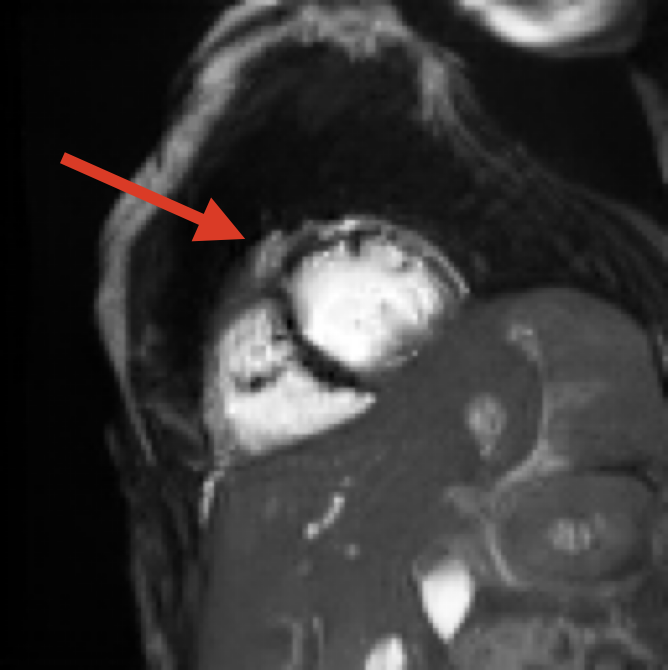}\label{fig:underweight}}\hfill
    \subfloat[Normal]{\includegraphics[width=0.20\linewidth]{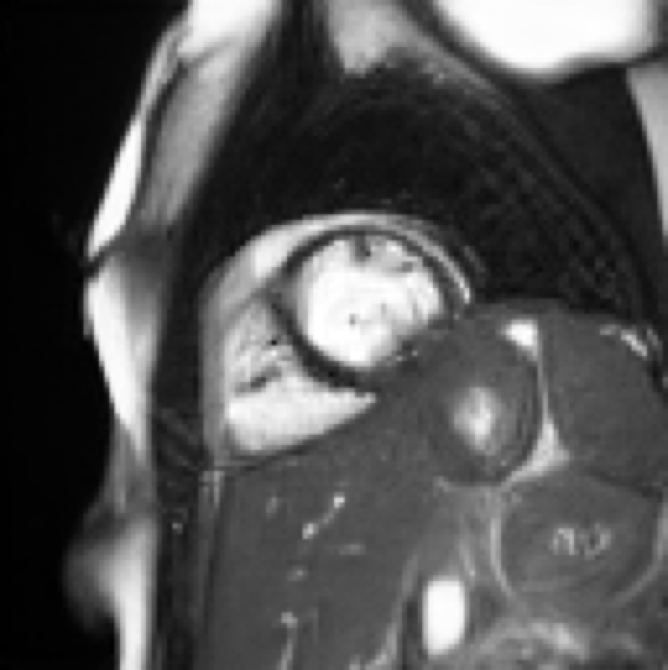}\label{fig:normal}}\hfill
    \subfloat[Overweight]{\includegraphics[width=0.20\linewidth]{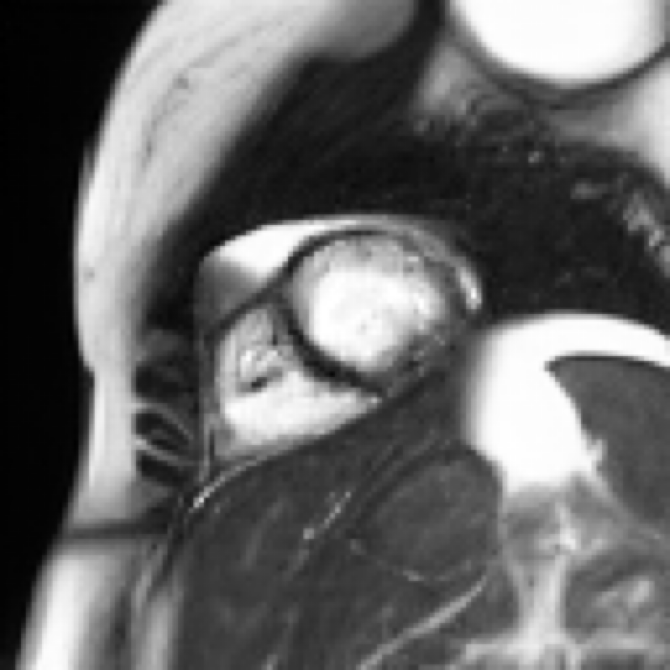}\label{fig:overweight}}\hfill
    \subfloat[Obese]{\includegraphics[width=0.20\linewidth]{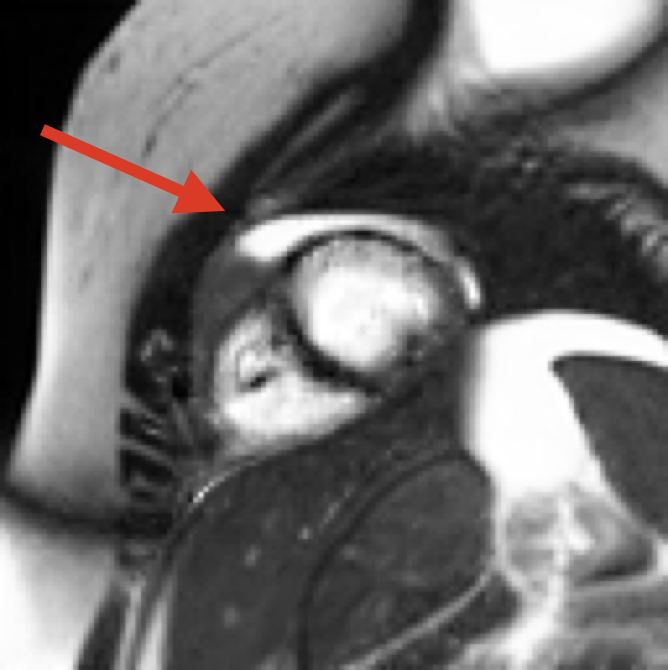}\label{fig:obese}}
    
    \caption{Effect of altering sensitive features on the generated images using the prompt: "Female, age in 60s, \textit{\{BMI category\}}". In this example, the sensitive attribute BMI was modified between generation runs to observe its effect on the generated CMR scans.}
    \label{fig:bmi_change}

\end{figure}

\begin{figure}[tb]
    \centering
    \subfloat[Real]{\includegraphics[width=0.20\linewidth]{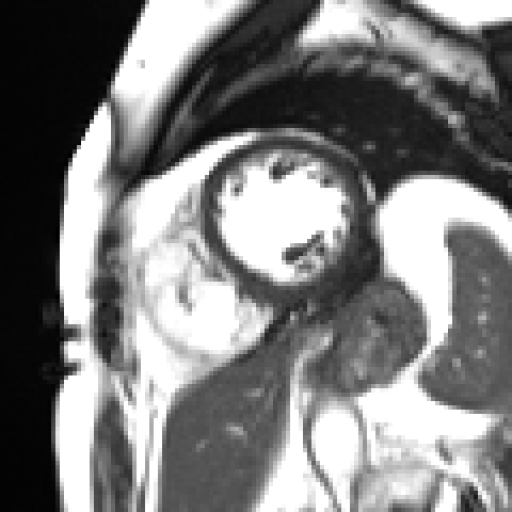}\label{fig:real_1}}\hfill
    \subfloat[Synthetic \#1]{\includegraphics[width=0.20\linewidth]{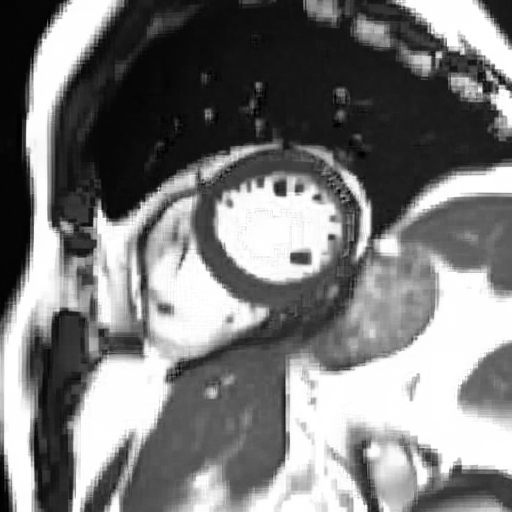}\label{fig:seed_5}}\hfill
    \subfloat[Synthetic \#2]{\includegraphics[width=0.20\linewidth]{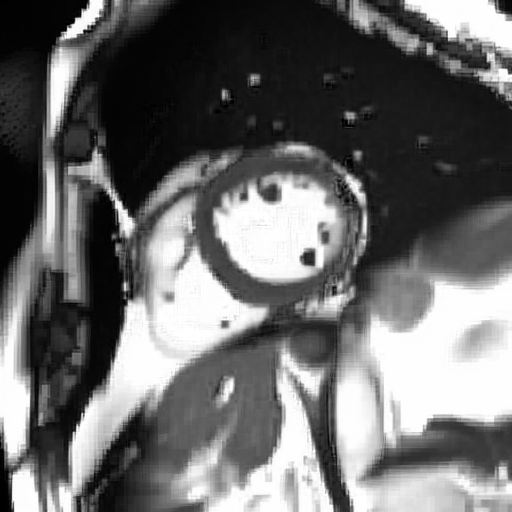}\label{fig:seed_2}}\hfill
    \subfloat[Synthetic \#3]{\includegraphics[width=0.20\linewidth]{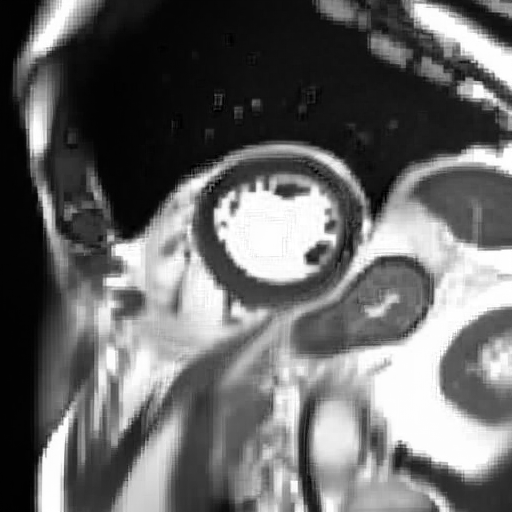}\label{fig:seed_3}}
    
    \caption{Variability in the generated CMR images achieved by using same input data, but different seeds for prompt: \textit{Female, age in 70s, overweight BMI, with heart failure}. \ref{fig:real_1} Reproduced by kind permission of UK Biobank ©.}
    \label{fig:seeds}

\end{figure}

Sample images in Fig.\ref{fig:bmi_change} demonstrate the model’s ability to link visual BMI indicators with textual prompts. Increased pericardial adipose tissue (PAT), marked with red arrows, is visible as BMI progresses. As noted in~\cite{szabo_radiomics_2023} PAT is a significant factor in discrimination of HF patients. CMR images generated with different seeds for the same input (Fig.~\ref{fig:seeds}) further showcase the model’s ability to create a diverse set of samples and highlight the augmentation potential of the proposed approach.

\subsection{Downstream task: Heart failure classification}

\begin{figure}[tb]
    \includegraphics[trim=7 7 7 7, clip, width=.27\linewidth]{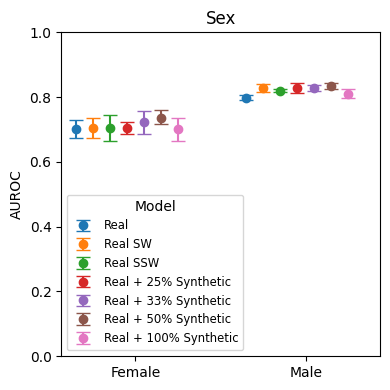}\hfill
    \includegraphics[trim=7 7 7 7, clip, width=.27\linewidth]{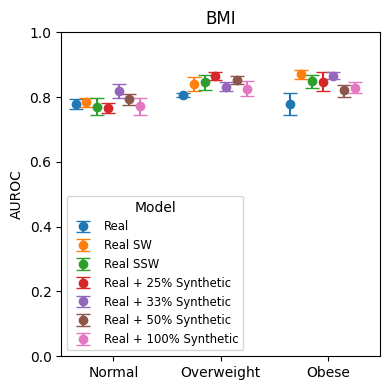}\hfill
    \includegraphics[trim=7 7 7 7, clip, width=.27\linewidth]{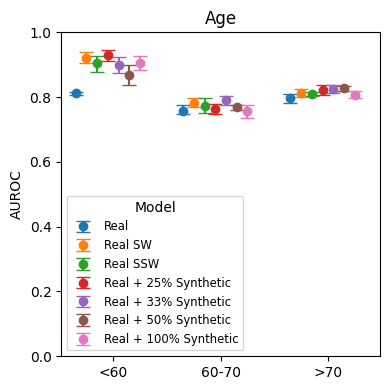}
    
    \caption{Mean AUROC with 95\% CI for each subgroup within the sensitive attributes.}
    \label{fig:auroc}
\end{figure}


\begin{table}[t]
\caption{Average of per-group cardiac disease classification (CLF) scores for each sensitive attribute and overall performance for the whole population. CLF$_{Real+Synth}$ uses 33\% synthetic data. Values multiplied by 100; best results in bold.}

\label{tab:auroc_eod}
\setlength{\tabcolsep}{14.5pt} 

\scriptsize

\begin{tabular}{l l|ccc}
\toprule
Group & Metric & CLF$_{Real SW}$ & CLF$_{Real SSW}$ & CLF$_{Real+Synth}$ \\
\midrule
\multirow{3}{*}{Sex} 
  & AUROC $\uparrow$   & 78.9$\pm$1.5  & 78.3$\pm$1.7  & \textbf{79.6$\pm$1.6} \\
  & BACC $\uparrow$    & 70.4$\pm$1.2  & 68.4$\pm$1.3  & \textbf{72.4$\pm$1.0} \\
  & EOD $\downarrow$   & 37.2$\pm$5.6  & 40.9$\pm$4.7  & \textbf{32.6$\pm$6.1} \\
\midrule
\multirow{3}{*}{BMI}
  & AUROC $\uparrow$   & 83.1$\pm$1.3  & 82.2$\pm$0.9  & \textbf{83.8$\pm$0.8} \\
  & BACC $\uparrow$    & 73.7$\pm$0.9  & 72.1$\pm$1.2  & \textbf{75.2$\pm$0.7} \\
  & EOD $\downarrow$   & 39.7$\pm$5.7  & 37.8$\pm$8.1  & \textbf{32.1$\pm$4.6} \\
\midrule
\multirow{3}{*}{Age}
  & AUROC $\uparrow$   & \textbf{83.7$\pm$1.0} & 82.8$\pm$0.9  & \textbf{83.7$\pm$0.9} \\
  & BACC $\uparrow$    & 74.9$\pm$1.2  & 72.9$\pm$1.5  & \textbf{76.2$\pm$0.8} \\
  & EOD $\downarrow$   & 20.8$\pm$6.1  & \textbf{17.2$\pm$3.6} & 23.7$\pm$7.5 \\
\midrule
\multirow{2}{*}{Overall}
  & AUROC $\uparrow$   & 83.1$\pm$1.1  & 82.4$\pm$0.9  & \textbf{83.6$\pm$0.8} \\
  & BACC $\uparrow$    & 74.3$\pm$0.9  & 72.7$\pm$1.2  & \textbf{75.8$\pm$0.7} \\
\bottomrule
\end{tabular}
\end{table}

As presented in Table~\ref{tab:auroc_eod}, integrating synthetic data with real samples led to an overall improvement in disease diagnosis performance, as reflected in higher AUROC and BACC scores, as well as improved average per-attribute scores. Specifically, the average BACC increased by 2\% for groups separated by sex, 1.5\% for BMI, and 1.3\% for age. The ablation study in Fig.~\ref{fig:auroc} further illustrates the impact of varying the proportion of synthetic data used during training. While the results exhibit a notable level of noise due to the limited number of test samples, a visible trend emerges -- combining real and synthetic data provides a performance boost. 
On another note, while label-based weighting helps during training, subgroup weighting does not provide additional boost, likely due to much smaller subgroup sizes causing overfitting.
From a fairness perspective, EOD improved for both sex and BMI attributes, though a slight decrease was observed for age. These findings are consistent with the synthetic data quality analysis presented in \ref{synth_evaluation}. As shown in Fig.~\ref{fig:auroc}, females and individuals with a normal BMI experienced the most significant performance gains, effectively narrowing the gap to better-performing groups (e.g., males and obese individuals) by 13\%. This aligns with the distribution imbalance observed in Fig.~\ref{fig:demographics}, where these populations had the lowest prevalence in the dataset, highlighting the potential of synthetic data to mitigate biases in model performance.

\section{Discussion and Conclusion}


In this work, we explore the use of generative latent diffusion models to address biases in CMR datasets. We show that combining textual (sex, age, BMI, heart condition) and imaging inputs (segmentation masks of cardiac shape) enables flexible and controllable synthetic data generation. Empirical evaluation on cardiac disease classification demonstrates performance gains for average per-group scores and fairness when training with synthetic balanced data, highlighting the potential of targeted data augmentation for reducing bias in cardiac imaging datasets. Our results also illustrate the challenge of addressing fairness in low-prevalence diseases, where subgroup sizes remain small and noisy even in large datasets. Future work on larger cohorts and more common diseases is needed to further assess this approach, including evaluation of subgroup-specific feature interpretability and analysis of changes in uncertainty estimates across subgroups. 
Additionally, we illustrate that such data augmentation is feasible on modest hardware, with all models---including multi-conditional diffusion models---trained on a single consumer-grade GPU, thereby laying the foundation for broader clinical adoption across diverse healthcare settings with varying computational resources.


    
\begin{credits}
\subsubsection{\ackname} 
This work was conducted using the UK Biobank resource under access application 2964. It received funding from the European Union’s Horizon Europe research and innovation programme, Grant Agreement No. 101057849 (DataTools4Heart). NA acknowledges support from MRC Clinician Scientist Fellowship (MR/X020924/1).
SEP declares consultancy for Circle Cardiovascular Imaging, Inc., Calgary, Canada. Barts Charity (G-002346) contributed to fees required to access UK Biobank data. This work acknowledges the support of the National Institute for Health and Care Research Barts Biomedical Research Centre (NIHR203330); a delivery partnership of Barts Health NHS Trust, Queen Mary University of London, St George’s University Hospitals NHS Foundation Trust and St George’s University of London.


\end{credits}

\bibliographystyle{splncs04}
\bibliography{Paper-0015}

\end{document}